# SUSEP-Net: Simulation-Supervised and Contrastive Learning-based Deep Neural Networks for Susceptibility Source Separation

Min Li[1], Chen Chen[1], Zhenghao Li[2], Yin Liu[3], Shanshan Shan[4], Peng Wu[5], Pengfei Rong[3], Feng Liu[6], G. Bruce Pike[7], Alan H.Wilman[8], Hongfu Sun[9], Yang Gao[1,*]

[1]School of Computer Science and Engineering, Central South University, Changsha, China

[2]School of Biomedical Engineering, Shanghai Jiao Tong University, Shanghai, China

[3]Department of Radiology, The Third Xiangya Hospital, Central South University, Changsha, China

[4]State Key Laboratory of Radiation, Medicine and Protection, School for Radiological and Interdisciplinary Sciences (RAD-X), Collaborative Innovation Center of Radiation Medicine of Jiangsu Higher Education Institutions, Soochow University, Suzhou, China.

[5]Philips Healthcare, Shanghai, China

[6]School of Electrical Engineering and Computer Science, University of Queensland, Brisbane, Australia

[7]Department of Radiology and Clinical Neurosciences, Hotchkiss Brain Institute, University of Calgary, Calgary, Canada.

[8]Department of Radiology and Diagnostic Imaging, University of Alberta, Edmonton, Canada

[9]School of Engineering, University of Newcastle, Newcastle, Australia.

*Correspondence: Yang Gao

Address: Room 429, Information Building,

Central South University, Changsha, China.

Email: yang.gao@csu.edu.cn

## Abstract

Quantitative susceptibility mapping (QSM) provides a valuable tool for quantifying susceptibility distributions in human brains; however, two types of opposing susceptibility sources (i.e., paramagnetic and diamagnetic), may coexist in a single voxel, and cancel each other out in net QSM images. Susceptibility source separation techniques enable the extraction of sub-voxel information from QSM maps. This study proposes a novel SUSEP-Net for susceptibility source separation by training a dual-branch U-net with a simulation-supervised training strategy. In addition, a contrastive learning framework is included to explicitly impose similarity-based constraints between the branch-specific guidance features in specially-designed encoders and the latent features in the decoders. Comprehensive experiments were carried out on both simulated and *in vivo* data, including healthy subjects and patients with pathological conditions, to compare SUSEP-Net with three state-of-the-art susceptibility source separation methods (i.e., APART-QSM, $\chi$-separation, and $\chi$-sepnet). SUSEP-Net consistently showed improved results compared with the other three methods, with better numerical metrics, improved high-intensity hemorrhage and calcification lesion contrasts, and reduced artifacts in brains with pathological conditions. In addition, experiments on an agarose gel phantom data were conducted to validate the accuracy and the generalization capability of SUSEP-Net.

## 1. INTRODUCTION

Quantitative Susceptibility Mapping (QSM) is a valuable MRI technique that can quantify tissue magnetic susceptibility distribution from MRI phase signals (Deistung et al., 2017; Liu et al., 2015; Wang & Liu, 2015). It has demonstrated great potential in studying various neurological diseases, e.g., Parkinson's Disease (PD) (Acosta-Cabronero et al., 2017; Langkammer et al., 2016), Alzheimer's Disease (AD) (Acosta-Cabronero et al., 2013; Ayton et al., 2017), Multiple Sclerosis (MS) (Elkady et al., 2018; Elkady et al., 2019; Langkammer et al., 2013; Li et al., 2016; Wisnieff et al., 2015), intracranial hemorrhage (ICH) (De et al., 2020; Sun et al., 2016; Sun et al., 2018), mild cognitive impairment (MCI) (L. Chen et al., 2021), as well as healthy brain aging (Bilgic et al., 2012; Zhang et al., 2018).

There are two major types of opposite susceptibility sources in human brains, e.g., paramagnetic (positive susceptibility) iron and diamagnetic (negative susceptibility) myelin, both playing important roles in the central nervous system. For instance, iron biochemistry and the disruption of iron homeostasis may be involved in PD pathogenesis (Guan et al., 2019; Zecca et al., 2004), while myelin's abnormal production and degradation are related to neurodegenerative diseases like MS (Compston & Coles, 2008) and leukodystrophy (Nave, 2010). These two types of opposing susceptibility sources can coexist within a single QSM voxel due to the limited acquisition resolution, cancelling each other out and resulting in inaccurate quantification and reducing QSM's specificity for individual substances within a single voxel. For example, it is challenging to use QSM to distinguish iron-amyloid β interactions in Alzheimer's disease (AD) (Becerril-Ortega et al., 2014; Wärmländer et al., 2019) or iron-myelin colocalization in MS lesions (Ji et al., 2024).

Different methods have been proposed for susceptibility source separation, i.e., separating QSM into paramagnetic ($\chi_{pos}$) and diamagnetic ($\chi_{neg}$) components (J. Chen et al., 2021; Emmerich et al., 2021; Kim et al., 2025; Li et al., 2023; Schweser et al., 2011; Shin et al., 2021; Zhang et al., 2022). Specifically, a linear method was first proposed (Schweser et al., 2011) based on R2* and susceptibility images, assuming they were linearly related to iron and myelin concentrations. The more recent χ-separation method (Shin et al., 2021) improved the specificity by jointly taking advantage of local field (or the frequency shift) data and R2'(=R2* - R2) relaxometry, which is assumed to be linearly dependent on the absolute susceptibility, and the coefficients could be determined by a spatially uniform magnitude decay kernel. This method has shown great potential in the diagnosis of multiple sclerosis versus neuromyelitis optica spectrum disorder (Kim et al., 2023). APART-QSM (Li et al., 2023) improved this scheme by adopting voxel-specific decay kernels, accounting for regional variations in susceptibility mixtures and has been successfully adopted to study epilepsy in a recent work (Z. Li et al., 2025). DECOMPOSE-QSM (J. Chen et al., 2021), constructed based on a three-pool theory model, can estimate both para- and diamagnetic sources without R2' images, however, unrealistic static dephasing assumptions might reduce its accuracy, particularly in anisotropic white matter (Duyn & Schenck, 2017). More recently, a deep learning solution was also proposed, i.e., χ-sepnet (Kim et al., 2025), by training a U-net on a large number of *in vivo* datasets acquired with multiple head orientations, to learn the mapping from the *in vivo* local field, R2', and QSM images (network inputs) to traditional χ-separation results (training labels), which might not be the ideal ground truth due to the reconstruction errors in traditional χ-separation method. In addition, the acquisition of such sophisticated datasets is a non-trivial task.

In this work, we propose a novel **SUSEP-Net** for susceptibility source separation (i.e., the χ separation or sub-voxel QSM) from local field, R2', and QSM images, by training a dual-branch U-net on a simulated dataset with purely synthetic pathological lesions, with one branch responsible for $\chi_{pos}$ reconstruction and the other for $\chi_{neg}$. In addition, a contrastive learning strategy is also introduced to impose similarity-based constraints between the branch-specific guidance features. Comprehensive experiments were conducted to compare SUSEP-Net with state-of-the-art χ-separation methods using simulated brain data and *in vivo* human brain subjects from healthy volunteers and patients with carbon monoxide (CO) poisoning and myelin oligodendrocyte glycoprotein antibody-associated disease (MOGAD). A phantom experiment was also conducted to validate the accuracy and generalization capability of the proposed method.

## 2. METHOD

### 2.1 Theoretical Model for Susceptibility Source Separation

The simplified model for the magnetic source separation problem can be described with the following complex equation(Li et al., 2023; Shin et al., 2021):

$$R2'(r) + i \cdot \Delta B_{local}(r) = \text{A}(\text{r}) \cdot \big( (\chi_{pos}(r) - \chi_{neg}(r)) \\ + i \cdot D(r) \otimes (\chi_{pos}(r) + \chi_{neg}(r)) \big), \tag{1}$$

where $r$ is the spatial coordinate, $R2'(r) = R2^* - R2$ is the reversible transverse relaxation rate, $\Delta B_{local}$ is the local field induced by the tissues (i.e., $\chi_{pos}(r) + \chi_{neg}(r)$), A(r) denotes the voxel-specific magnitude decay kernel relating R2' and absolute susceptibility (i.e., $\chi_{pos}(r) - \chi_{neg}(r)$), $D(r)$ represents the unit dipole kernel, and $\otimes$ denotes spatial convolution. Equation (1) is also adopted as the simulated training data generator in this work.

## 2.2 SUSEP-Net Architecture

As shown in Fig. 1(a), the proposed SUSEP-Net is composed of three stages: (1) Feature Extraction (FE), (2) Feature Integration (FI), and (3) Feature Aggregation (FA). It is mainly constructed from a dual-branch U-net with one shared encoder (the orange blocks) and two decoders (i.e., the top and bottom network modules in FI and FS), with the top branch for paramagnetic component ($\chi_{pos}$) reconstruction and the bottom branch for $\chi_{neg}$ reconstruction. In addition to the primary backbone, two additional encoders (i.e., Encoder$_{pos}$ and Encoder$_{neg}$ in the dashed boxes in FE stage) were incorporated to generate two guidance features for $\chi_{pos}$ and $\chi_{neg}$ reconstruction tasks, respectively, and in the meantime, enable the proposed contrastive learning. More details about the SUSEP-Net and the contrastive learning are described as follows:

**Feature Extraction:** The FE part of the proposed SUSEP-Net mainly comprises three independent encoders, which are composed of three consecutive encoding blocks, as commonly used in traditional U-net encoding part, gradually increasing the feature numbers from 64 to 256 and decreasing the spatial resolution of hidden features for multi-scale learning. In this work, each encoding block is composed of two convolutional layers (kernel size: 3×3×3 and stride size: 1×1×1), two batch-normalization layers, and one max-pooling (kernel size: 2×2×2 and stride size: 2×2×2).

Suppose that the inputs (i.e., R2', local field, and initial QSM estimation) to SUSEP-Net are represented as $R2' \in \mathbb{R}^{N_x \times N_y \times N_z}$, $\delta B \in \mathbb{R}^{N_x \times N_y \times N_z}$, and $\chi \in \mathbb{R}^{N_x \times N_y \times N_z}$, respectively, where $N_x$, $N_y$, and $N_z$ are the image height, width and depth, then the operations in FE stage can be mathematically denoted as:

$$F_v = Enc_1(X_{in})$$

$$Guide_{pos} = Enc_2(QSM),$$

$$Guide_{neg} = Enc_3(QSM)$$

(2)

where $X_{in} = Concat(R2', \delta B, \chi) \in \mathbb{R}^{3 \times N_x \times N_y \times N_z}$ is the concatenation of the three inputs along the channel dimension, and $Enc_{1,2,3}$ represent the above-mentioned encoders, which are basically the same except that the input channel number is set as 3 for $Enc_1$ and 1 for the other two. $F_v$, $Guide_{pos}$, and $Guide_{neg}$ denote the latent and guidance features in Fig. 1(a).

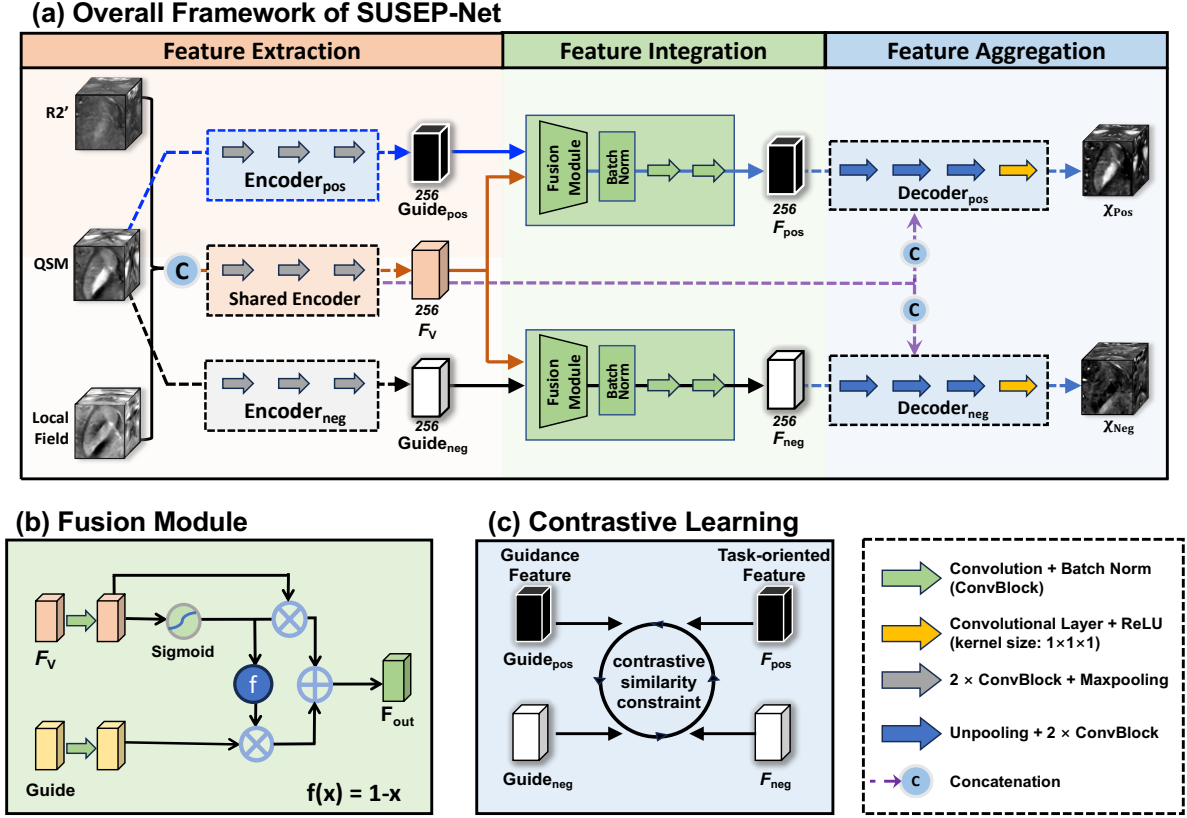

**Figure 1.** Overall framework of the proposed (a) SUSEP-Net, which is developed based on a dual-branch U-net backbone and two additional encoders (i.e., Encoder_pos and Encoder_neg in the dashed boxes) for contrastive learning, taking the R2', local field, and QSM images (for generating the guidance features) as inputs and producing the χ separation images (χ_pos and χ_neg) as outputs. The numbers under the feature cubes represent the channel number of the corresponding hidden feature. (b) illustrates the fusion module in the feature interaction part in (a), while (c) shows the diagram of contrastive learning, which is designed to enhance mutual interactions of features in both the top and bottom branches of SUSEP-Net.

**Feature Integration:** At the FI stage, SUSEP-Net tries to fuse the features produced in FE stage to form two task-oriented features (i.e., $F_{pos}$ and $F_{neg}$ in Fig. 1(a)) for the final image reconstruction task, using a fusion module (Fig. 1(b)) developed in our recent work (M. Li et al., 2025), followed by a batch normalization layer and two consecutive convolutional blocks (kernel size: 3×3×3 and stride size: 1×1×1). The fusion module can be mathematically described as follows:

$$\alpha = \text{sigmoid}\left(Conv3D\left(Guide_{1,2}\right)\right)$$
$$F_{out}^{1,2} = Conv3D\left(Guide_{1,2}\right) * \alpha + Conv3D(F_v) * (1-\alpha) \tag{3}$$

where $Guide_{1,2} = Guide_{pos,neg}$ are the guidance features generated in FE stage, and $F_{out}^{1,2}$ will be used for the generation of task-oriented $F_{pos}$ and $F_{neg}$ correspondingly, which will be used for the following image recovery part and contrastive learning.

**Feature Aggregation:** The FA stage is composed of two independent U-net decoders that are responsible for $\chi_{pos}$ and $\chi_{neg}$ reconstructions, respectively. Each decoder comprising three decoding blocks, containing two convolutional layers (kernel size: 3×3×3 and stride size: 1×1×1), two batch normalization layers, and two unpooling layers (transposed convolutions with kernel size: 2×2×2 and stride size: 2×2×2). Similar to conventional U-nets (Ronneberger et al., 2015), features of the same scale from the intermediate features of $Enc_1$ were skip connected to the corresponding features in the decoders (i.e., concatenated along the channel dimension), to fully exploit the multi-scale hidden features. Two 1×1×1 convolution layers were adopted to aggregate all the latent features output from the final decoding block to generate the final $\chi_{pos}$ and $\chi_{neg}$ images.

### 2.2.2 Contrastive Learning

Inspired by several recent works investigating the foundational and debiased contrastive representation learning (Chen et al., 2020; Chuang et al., 2020; Cole et al., 2022; Oord et al., 2018) and prompt learning (Ge et al., 2023; Zhou et al., 2022a; Zhou et al., 2022b), we propose a contrastive learning strategy to explicitly build semantic links between the guidance vectors ($Guide_{pos, neg}$, i.e., the outputs of $Encoder_{pos}$ and $Encoder_{neg}$ in Fig. 1) and task-oriented features ($F_{pos, neg}$). A contrastive loss, i.e., Eq. (4) is incorporated during network training to maximize the similarity between the Guide-$F$ features in the same branch, and minimize the similarity between guidance and task-oriented features in different branches. The proposed coxntrastive learning strategy are defined as follows:

$$Loss_{Contrast} = -ln\left(\frac{exp(s(Guide_{pos}, F_{pos}))}{exp(s(Guide_{pos}, F_{pos})) + exp(s(Guide_{pos}, F_{neg}))}\right)$$
$$-ln\left(\frac{exp(s(Guide_{neg}, F_{neg}))}{exp(s(Guide_{neg}, F_{neg})) + exp(s(Guide_{neg}, F_{pos}))}\right) \tag{4}$$

where $s(x, y) = \left| \frac{<x,y>}{\|x\|_2^2 \cdot \|y\|_2^2} \right|$ represents a normalized similarity measure for vectors $x$ and $y$, $< x, y >$ is the inner product of vectors $x$ and $y$.

## 2.3 Training Datasets

Institutional ethics board approvals have been obtained for all MRI brain data used in this work. Similar to the acquisition settings in APART-QSM (Li et al., 2023) and χ-separation (Shin et al., 2021), the training datasets (32 subjects) were acquired at 3T (Phillips, Ingenia Elition X) with a 3D multi-echo GRE (mGRE) sequence with parameters: 8 unipolar echoes, first TE / ΔTE / TR = 4.1 / 4.0 / 35 ms; matrix size= 224 × 224 × 128; voxel size = 0.9 mm × 0.9 mm ×1 mm; flip angle = 20°; CS-SENSE acceleration factor = 3; total acquisition time = 5.43 mins, and a 2D GraSE sequence with parameters: 5 echoes, effective TE / ΔTE / TR = 20 / 20 / 3000 ms; matrix size= 196 × 196; in-plane voxel size = 1 mm ×1 mm; slice thickness = 1 mm, 128 slices; with EPI factor =5, TSE factor = 5, and SENSE acceleration factor = 2, total scan time = 6.6 mins. The APART-QSM (Li et al., 2023) reconstruction pipeline was conducted to obtain the corresponding full-sized $\chi_{pos}$, $\chi_{neg}$, and voxel-specific magnitude decay kernel (i.e., A(r) in Eq. (1) and the A map in Fig. 2), which serves the training labels for SUSEP-Net.

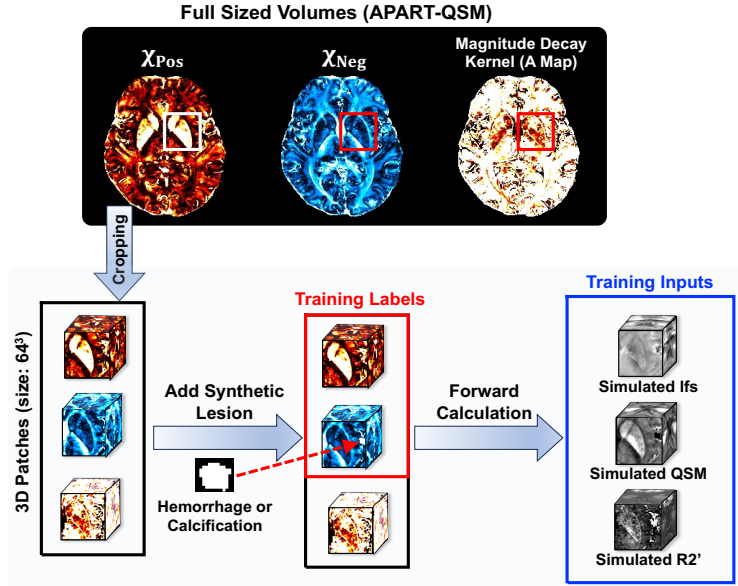

**Figure 2.** Training data simulation framework, including the reconstruction of full-size volumes ($\chi_{pos}$, $\chi_{neg}$, and the magnitude decay kernel) using APART-QSM, generation of patches ($\chi_{pos}$ and $\chi_{neg}$) cropped from the full-size volumes as training patches, and an additional pure synthetic source generated with simple geometric shapes, then the forward calculation model for susceptibility source separation to simulate training inputs (i.e., simulated local field, R2' and QSM patches).

Figure 2 shows the simulation-supervised training framework adopted in this work for the training of SUSEP-Net, similar to our recently proposed iQSM (Gao et al., 2022) and iQSM+ (Gao et al., 2024), based on 3024 paired training patches (size: 64×64×64) cropped from 36 full-size $\chi_{pos}$ and $\chi_{neg}$ volumes (size: 224×224×128) using a sliding window of size $64^3$ with stride of 24×36×20. In addition to the "realistic" positive and negative susceptibility sources cropped from the APART-QSM reconstructions of the 32 human brains, we also incorporated the pure-synthetic hemorrhage and calcifications lesions in the training dataset using the method and parameter proposed in iQSM (Gao et al., 2022), which indeed doubles the size of the training dataset. Constant susceptibility was randomly assigned from uniform distributions with ranges of [0.4, 1.2] and [−0.3, −0.1] for the synthetic hemorrhage and calcification sources. Simulated hemorrhage sources were added into $\chi_{pos}$ training labels, while calcification lesions were incorporated into the $\chi_{neg}$ data. After obtaining the $\chi_{pos}$, $\chi_{neg}$, and the A map patches, all the network training inputs, i.e., simulated local field maps, R2', and the QSM patches, can be synthesized using Eq. (1).

## 2.4 Network Training

In this work, all network parameters were initialized with random numbers from a Gaussian distribution (zero mean, 0.01 standard deviation). Networks were trained for 100 epochs (14 hours) on an Nvidia Tesla A6000 GPU using the Adam optimizer, with learning rates being $10^{-3}$ for the first 30 epochs, $10^{-4}$ for epochs 30-60, and $10^{-5}$ for the final 40 epochs, using a composite loss functions defined as follows:

$$Loss = \alpha * Loss_{Contrast} + \beta * Loss_{l2} + \gamma * Loss_{model} + \delta * Loss_{gradient}, \quad (5)$$

where $\alpha = 1$, $\beta = 1$, $\gamma = 0.5$, and $\delta = 0.1$ are the empirical weighting parameters for the four different loss components. The $Loss_{Contrast}$ is defined as in Eq. (4) to increase the similarity between paired *Guide-F* features in the same branch and decrease the similarity between paired *Guide-F* features in different branches. The $Loss_{l2}$ is defined as Eq. (6) denoting the distance between network reconstructions and the training labels:

$$Loss_{l2} = MSE(\chi^*_{pos}, \text{GT}_{pos}) + MSE(\chi^*_{neg}, \text{GT}_{neg}), \quad (6)$$

where $\chi^*_{pos}$ and $\chi^*_{pos}$ are network reconstructions, while $\text{GT}_{pos}$ and $\text{GT}_{neg}$ denote the training labels.

The model loss $Loss_{model}$ is similar to the ones used in QSMnet (Yoon et al., 2018):

$$Loss_{model} = Loss_{QSM} + Loss_{lfs} + Loss_{R'_2}, \qquad (7)$$

where

$$\begin{cases} Loss_{QSM} = ||(\chi^*_{pos} - \chi^*_{neg}) - \chi|| \\ Loss_{lfs} = ||D * (\chi^*_{pos} - \chi^*_{neg}) - \Delta B_{local}||, \\ Loss_{R'_2} = ||A * (\chi^*_{pos} + \chi^*_{neg}) - R'_2|| \end{cases} \qquad (8)$$

where $\chi$, $\Delta B_{local}$, and $R'_2$ are the input QSM, local field, and R2' images; $D$ and $A$ are the dipole kernel and magnitude decay kernel in Eq .(1).

The final gradient loss is adopted to preserve edge information in the reconstructed map:

$$\begin{aligned} Loss_{gradient} = \ & || |\nabla\chi^*_{1,2}|_x - |\nabla GT_{1,2}|_x || + || |\nabla\chi^*_{1,2}|_y \\ & - |\nabla GT_{1,2}|_y || + || |\nabla\chi^*_{1,2}|_z - |\nabla GT_{1,2}|_z || \end{aligned}, \qquad (9)$$

where $\chi^*_{1,2}$ equals $\chi^*_{pos}$ and $\chi^*_{pos}$, respectively, and $GT_{1,2}$ equals $GT_{pos}$ and $GT_{neg}$, respectively.

In addition, all input data QSM, R2', LFS) were normalized as (data - mean) / std, where mean and std denote dataset-wise mean and standard deviation values. The pre-trained SUSEP-Net network and source codes are available at: https://github.com/YangGaoUQ/SUSEP-Net

## 3. EXPERIMENTS

### 3.1 Evaluation Datasets

To demonstrate the performance of the proposed SUSEP-Net with state-of-the-art susceptibility source separation methods, comprehensive experiments were conducted on the following datasets:

(1) Ten healthy brains synthesized using the pipeline in Fig. 2 were tested in an ablation to investigate the effectiveness of the proposed contrastive learning strategy, and to perform a quantitative comparsion of different susceptibility source separation algorithms. In addition, this dataset was also used to quantitatively compare deep grey matter (DGM) susceptibility measurements of different QSM methods.

(2) A simulated pathological brain with a hemorrhage source (1 ppm) and calcification source (-0.2 ppm) was superimposed onto one of the 10 healthy subjects in (1) to compare the performances of various algorithms on high-intensity susceptibility sources.

(3) An agarose gel phantom data scanned at 3T (uMR790, Unite Image Healthcare, Shanghai, China) was adopted to quantitatively validate the accuracy and the generalization capability of the proposed SUSEP-Net. The phantom is composed of nine cylindrical containers (arranged in 3 rows by 3 columns), with three cylinders in the first row containing single diamagnetic $CaCO_3$ solutions of increasing concentrations (i.e., 58.0, 116.0, and 174.0 mg/ml), the second row containing paramagnetic $Fe_3O_4$ solutions (i.e., 2.0, 4.0, and 6.0 ug/ml, respectively), and the third row containing the mixture of the previous two rows. More details about the phantom fabrication and scan parameters can be found in Fig. 1 of Ref (Li et al., 2023).

(4) Two pathological patients (one with carbon monoxide (CO) poisoning and one with myelin oligodendrocyte glycoprotein antibody-associated disease (MOGAD)) were scanned at 3T (Phillips, Ingenia Elition X), with the parameters described in the above Section 2.3, to qualitatively visualize and delineate lesions in susceptibility maps.

### 3.2 Reconstruction Pipeline and Performance Evaluation

For *in vivo* subjects, the local field maps and the QSM images were reconstructed directly from the acquired ME-GRE phase data using our recently proposed iQFM (Gao et al., 2022) and iQSM+ (Gao et al., 2024), respectively. The R2' images were calculated from R2* and R2 images, which were obtained through a mono-exponential fitting function in STI-Suite based on the mGRE and mSE magnitude images, respectively.

The proposed SUSEP-Net was compared with several state-of-the-art susceptibility source separation methods, including iterative χ-separation (Shin et al., 2021), APART-QSM (Li et al., 2023), and U-net-based χ-sepnet (Kim et al., 2025), on the above-described evaluation data. In addition to the χ-sepnet model provided by the original paper, we also trained another χ-sepnet model using the proposed simulation dataset to validate the effectiveness of the proposed synthetic training strategy using the phantom data. All deep learning inferences were conducted on one Nvidia A6000 GPU, and the traditional algorithms were finished on an Intel(R) Core(TM) i7-13700F CPU.

For quantitative comparison of various susceptibility source separation methods, the commonly used numerical metrics, i.e., Normalized Root Mean Squared Error (NRMSE) (Milovic et al., 2020), High-Frequency Error Norm (HFEN) (Milovic et al., 2020), and XSIM (Milovic et al., 2025) were calculated for simulation studies. In addition, susceptibility measurements of one frontal white matter (FWM) region, and seven deep grey matter regions, including globus pallidus (GP), putamen (PU), caudate (CN), substantia nigra (SN), red nucleus (RN), internal capsule (IC), thalamus (TH), with region-of-interests (ROIs) drawn manually using ImageJ (National Institutes of Health, Bethesda, MD), were measured to facilicate a quantitative comparision of different methods. Moreover, susceptibility line profiles crossing the simulated paramagnetic hemorrhage and diamagnetic calcification sources were also compared. Linear regressions of the mean susceptibility measurements against the solution concentration and the single-source susceptibility versus dual-source susceptibilities reconstructed from the mixture were also investigated using the agarose gel phantom, based on the circular ROIs (i.e., the nine cylindrical containers) manually drawn in 9 consecutive slices.

## 4. RESULT

### 4.1 Ablation Study: The Effectiveness of Contrastive Learning

An ablation study was conducted in this work to validate the effectiveness of the proposed contrastive learning strategy, and the quantitative results on the 10 simulated healthy brains are reported in Table 1. The proposed contrastive learning resulted in on average 3.2%, 2.9%, and 1.2% improvements in NMRSE, HFEN, and XSIM metrics, implying its effectiveness for disentangling paramagnetic and diamagnetic magnetic susceptibility sources.

**Table 1**. Ablation study of Contrastive Learning (CL) in SUSEP-Net based on 10 simulated healthy brains

|  | Method | NRMSE(%) | HFEN(%) | XSIM(%) |
|---|---|---|---|---|
|  |  |  | mean ± std |  |
| $\chi^{Pos}$ | SUSEP-Net | **5.08 ± 0.53** | **4.71 ± 0.64** | **98.70 ± 0.07** |
|  | SUSEP-Net w/o CL | 8.71 ± 0.83 | 8.14 ± 0.76 | 97.04 ± 0.55 |
| $\chi^{Neg}$ | SUSEP-Net | **10.44 ± 0.85** | **9.57 ± 0.70** | **98.21 ± 0.39** |
|  | SUSEP-Net w/o CL | 13.12 ± 0.92 | 11.91 ± 0.84 | 97.47 ± 0.46 |

### 4.2 Simulated Dataset

Figure 3 compares different susceptibility source separation methods on a simulated pathological brain with synthesized paramagnetic hemorrhage (1 ppm) and diamagnetic calcification (0.2 ppm) lesions. The proposed SUSEP-Net achieved the best numerical metrics for both $\chi_{Pos}$ and $\chi_{Neg}$ images. For example, in $\chi$-neg, the NRMSE of SUSEP-Net is 9.66%, compared to 24.72%, 31.97%, and 15.85% of APART-QSM, $\chi$-separation, and $\chi$-sepnet, respectively. In addition, SUSEP-Net also achieved the most accurate hemorrhage (0.97 $\pm$ 0.018 ppm) and calcification (0.19 $\pm$ 0.017 ppm) measurements, compared to 0.87 $\pm$ 0.023 ppm and 0.16 $\pm$ 0.030 ppm from APART-QSM, 1.15 $\pm$ 0.029 ppm and 0.14 $\pm$ 0.042 ppm from $\chi$-separation, 1.06 $\pm$ 0.024 ppm and 0.17 $\pm$ 0.021 ppm from $\chi$-sepnet. Line profiles of SUSEP-Net, APART-QSM, $\chi$-separation, and $\chi$-sepnet crossing the simulated lesions on the simulated pathological brain were plotted in Fig. 3(b), which further confirmed that the proposed SUSEP-Net achieved the closest measurements compared with the ground truth.

**Table 2**. Comparison of SUSEP-Net with APART-QSM, $\chi$-separation, and $\chi$-sepnet on 10 simulated healthy brains

|  | Method | NRMSE(%) | HFEN(%) | XSIM(%) |
|---|---|---|---|---|
|  |  |  | mean $\pm$ std |  |
| $\chi_{Pos}$ | APART-QSM | 11.52 $\pm$ 1.80 | 13.11 $\pm$ 2.26 | 96.57 $\pm$ 0.95 |
|  | $\chi$-separation | 16.41 $\pm$ 0.68 | 17.60 $\pm$ 1.50 | 94.28 $\pm$ 1.42 |
|  | $\chi$-sepnet | 10.27 $\pm$ 0.75 | 10.23 $\pm$ 0.68 | 98.18 $\pm$ 0.53 |
|  | **SUSEP-Net** | **5.08 $\pm$ 0.53** | **4.71 $\pm$ 0.64** | **98.70 $\pm$ 0.07** |
| $\chi_{Neg}$ | APART-QSM | 15.75 $\pm$ 1.48 | 12.95 $\pm$ 1.41 | 97.83 $\pm$ 0.81 |
|  | $\chi$-separation | 22.52 $\pm$ 0.70 | 21.95 $\pm$ 1.26 | 92.74 $\pm$ 1.18 |
|  | $\chi$-sepnet | 15.37 $\pm$ 0.91 | 14.93 $\pm$ 0.57 | 97.65 $\pm$ 0.29 |
|  | **SUSEP-Net** | **10.70 $\pm$ 0.96** | **9.55 $\pm$ 0.84** | **98.07 $\pm$ 0.46** |

In addition, the proposed SUSEP-Net was also compared with the other susceptibility source separation methods on the 10 healthy simulated brains, with NRMSE, HFEN, and XSIM reported in Table 3. The proposed SUSEP-Net, on average, achieved the best numerical metrics compared with the other methods. For example, SUSEP-Net decreased the NRMSE to only 5.08% on the $\chi_{pos}$ images from 11.52%, 16.41%, and 10.27% using $\chi$-separation, APART-QSM, and $\chi$-sepnet methods, respectively.

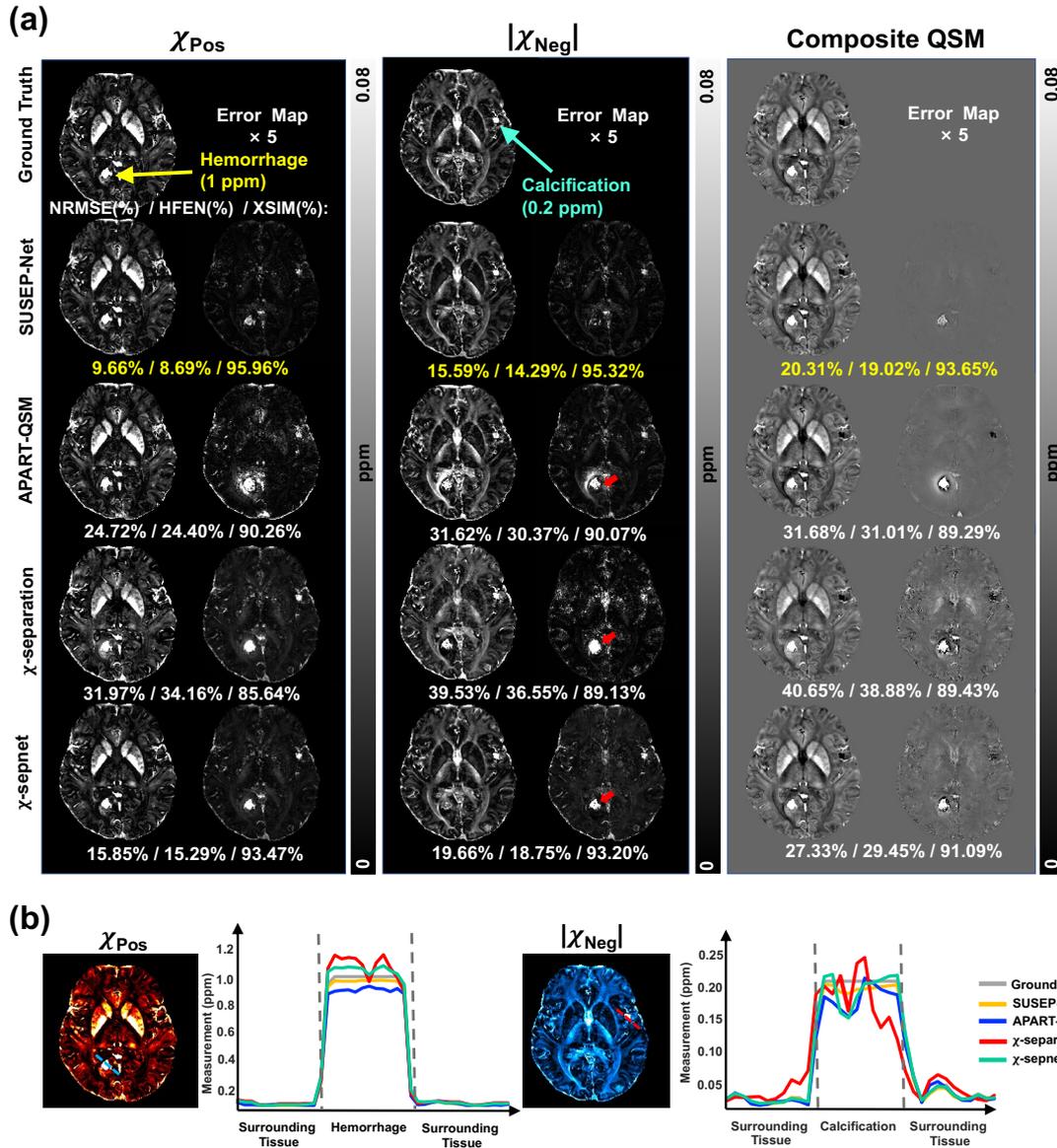

**Figure 3.** Comparison of the proposed SUSEP-Net with APART-QSM, χ-separation, and χ-sepnet on one simulated pathological brain. (a) shows the reconstruction results of different methods and 5× error maps relative to the simulated ground truth. The corresponding numerical metrics are reported below the images, with the best NRMSEs, HFENs, and XSIMs highlighted in yellow. Red arrows indicate visible reconstruction errors in $\chi_{neg}$ images around the hemorrhage lesions. (b) illustrates the susceptibility line profiles of all susceptibility source separation methods crossing the two simulated lesions.

Scatter plots of susceptibility measurements in five regions (i.e., GP, SN, TH, IC, and FWM) from the $\chi_{pos}$ and $\chi_{neg}$ maps of different methods on a simulated healthy subject are compared in Fig. 4 using linear regression. The proposed SUSEP-Net on average achieved the most accurate correlations to the ground truth, e.g., it resulted in a fitting slope of 0.95 ($R^2 = 0.87$) in the GP region, compared to 0.82 ($R^2$ = 0.83), 0.88 ($R^2$ = 0.72), and 0.89 ($R^2$ = 0.85) using APART-QSM, χ-separation, , and χ-sepnet, respectively.

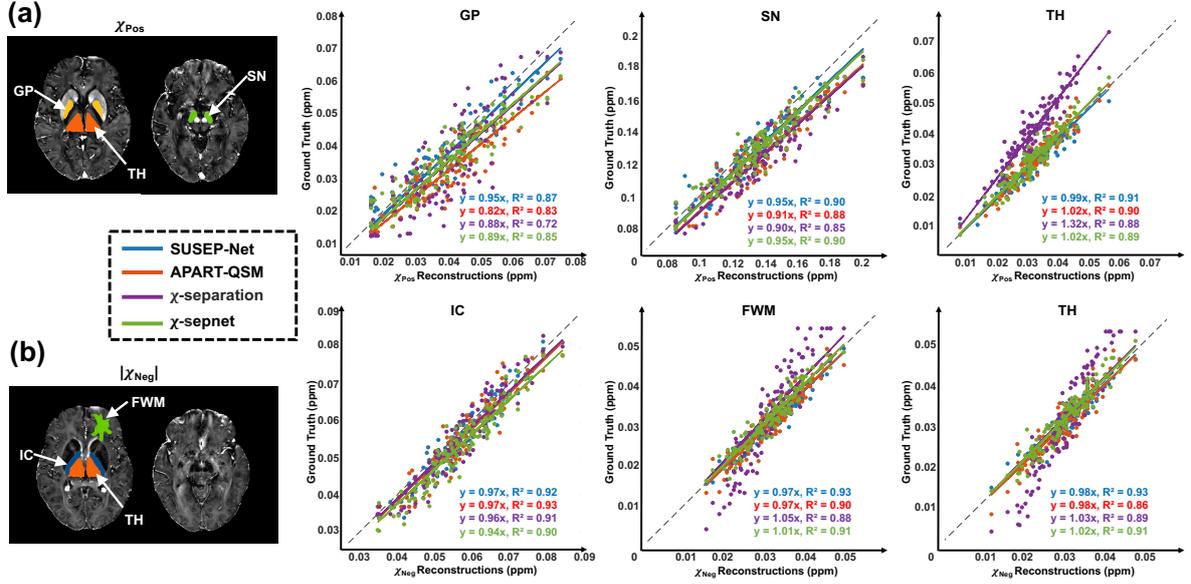

**Figure 4.** Comparison of linear regression analysis of susceptibility measurements in six brain regions from different susceptibility source separation methods. (a) shows the results in globus pallidus (GP), substantia nigra (SN), and thalamus (TH) from $\chi_{pos}$ reconstructions, while (b) demonstrates corresponding scatter plots for internal capsule (IC), thalamus (TH), and frontal white matter (FWM) of $\chi_{neg}$ results.

### 4.3 Agarose Gel Phantom Validation

To test the generalization capability and validate the accuracy of the proposed SUSEP-Net, the source separation results of SUSEP-Net were performed on a specifically designed phantom(described in Section 3.1(3)), and the results are compared with APART-QSM, χ-sepnet, and χ-sepnet (retrained using the same data as SUSEP-Net) in Fig. 5. All the deep learning methods achieved visually reasonable results for this phantom with no apparent artifacts. Specifically, SUSEP-Net exhibited the best linear correlations ($R^2$ = 0.96 for $CaCO_3$ and $R^2$ = 0.97 for $Fe_3O_4$) between the susceptibility measurements and the solution concentrations, as shown in Fig. 5(c). In addition, the SUSEP-Net was the only deep learning method that achieved nearly the same performances (slope = 0.96 and $R^2$ = 0.99) as AFTER-QSM on the linear regression analysis between susceptibility values from the single-source cylinders and those from mixed-source cylinders in Fig. 5 (d). Additionally, the χ-sepnet (retrained) consistently demonstrated improved results compared with the original χ-sepnet in Fig. (c) and (d) with better correlations.

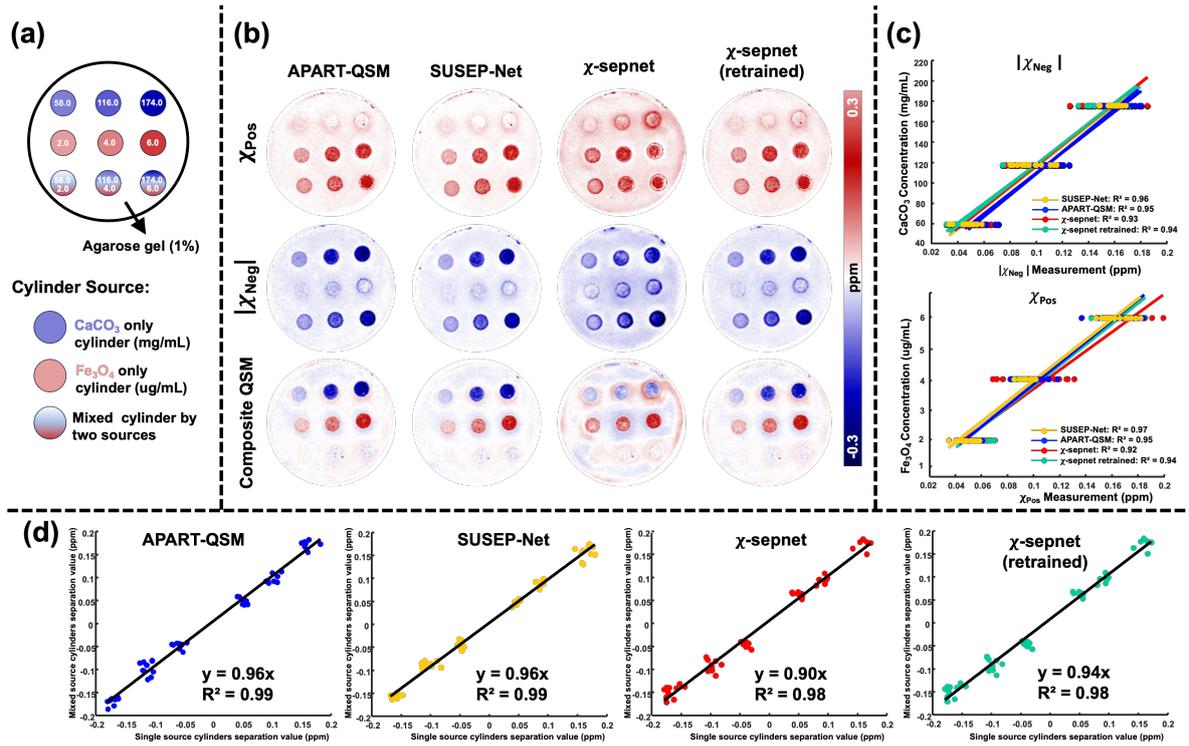

**Figure 5.** Validation of the accuracy and generalization capability of SUSEP-Net using a specifically designed phantom. (a) illustrates the diagram of the phantom compositions, (b) compares the reconstruction results of iterative APART-QSM and the deep learning based SUSEP-Net, χ-sepnet, and χ-sepnet (retrained) methods. (c) shows the linear regression results of the susceptibility values of different susceptibility source separation results against the solution concentrations. (d) compares the linear regressions of susceptibility measurements in the single-source cylinders (in the top two rows) versus the mixed-source cylinders (in the last row).

## 4.4 In Vivo Dataset

### 4.4.1 Healthy Subjects

Figure 6 compares the proposed SUSEP-Net with with χ-separation, APART-QSM, and χ-sepnet on one *in vivo* healthy brain. Overall, all methods led to promising results with similar contrast and great fine details as illustrated in the zoomed-in images. APART-QSM led to more artifacts in the slice 52 of $\chi_{neg}$ image near the sinus region. In addition, χ-separation and χ-sepnet might have produced more residual artifacts in the GP region on the $\chi_{neg}$ reconstructions, as pointed out by the red arrows in the zoomed-in images.

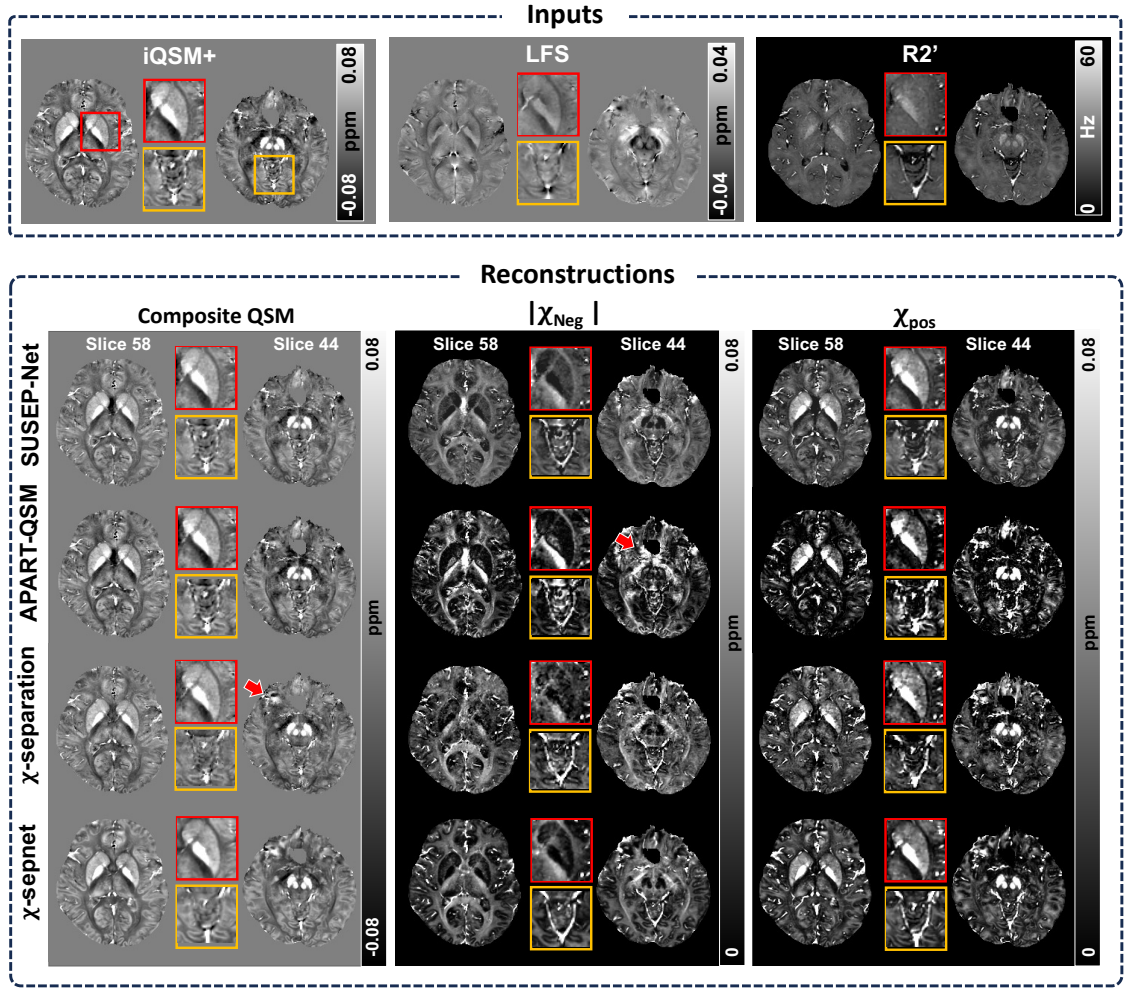

**Figure 6.** Comparison of different susceptibility separation methods on an *in vivo* brain acquired at 3T. The red arrows pointed to the potential artifacts in APART-QSM and χ-sepnet results.

Bar graphs in Fig. 7 compare different susceptibility source separation algorithms on the 10 healthy volunteers acquired at 3 T. The plots in the top panel compare DGM measurements of various methods on the paramagnetic maps ($\chi_{pos}$), while the bottom panel shows the bra graphs for diamagnetic maps ($\chi_{neg}$). Overall, for the $\chi_{pos}$-based comparison, the proposed SUSEP-Net showed similar measurements as the traditional APART-QSM and differed more from χ-separation and χ-sepnet results. For example, in the GP region, the mean susceptibility measurement of SUSEP-Net is 0.119 ppm, which is 0.027 ($P < 0.01$) and 0.011 ($P = 0.02$) lower than χ-separation and χ-sepnet, respectively. While for the $\chi_{neg}$ results, the proposed SUSEP-Net showed similar results to APART-QSM in GP (0.013 v.s. 0.015, $P = 0.54$) and TH (0.045 v.s. 0.044, $P = 0.61$), and are significantly lower than other methods in PU and SN.

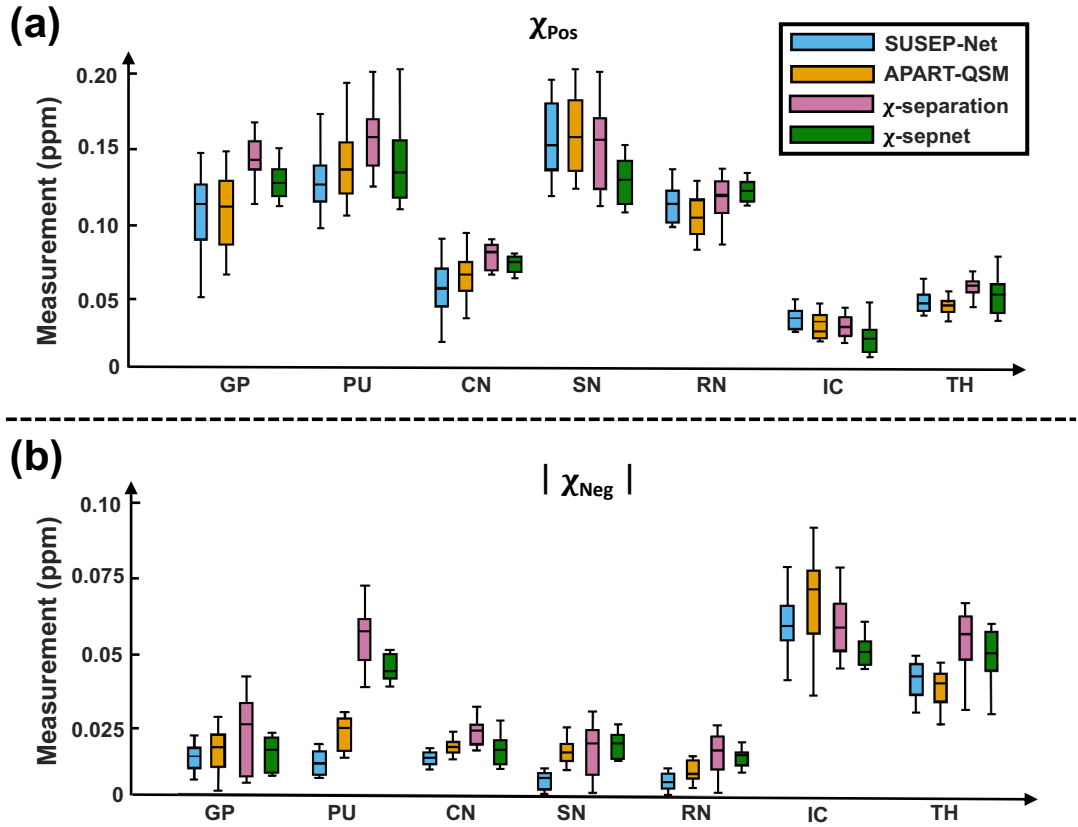

**Figure 7.** ROI analysis of different QSM separation methods on seven deep grey matter regions from 10 *in vivo* subjects at 3T, with (a) comparing the susceptibility of various methods on the paramagnetic components, and (b) illustrating the situations for diamagnetic components.

### 4.4.2 Pathological Brains

Figure 8 compares the susceptibility source separation images from two patients (one with CO poisoning and one with MOGAD) with apparent brain lesions. As shown in the reconstructed images, all methods successfully detected the lesions that appeared in R2' and the QSM images. Overall, deep learning-based SUSEP-Net and χ-sepnet showed better image contrasts than traditional APART-QSM and χ-separations. Specifically, purple arrows in Fig. 8 point to the potential reconstruction artifacts in conventional algorithms, which are absent in SUSEP-Net and χ-sepnet. Furthermore, the proposed SUSEP-Net showed more visually obvious abnormalities among all methods (green arrows), which may make it better for detecting abnormal lesions in pathological patients.

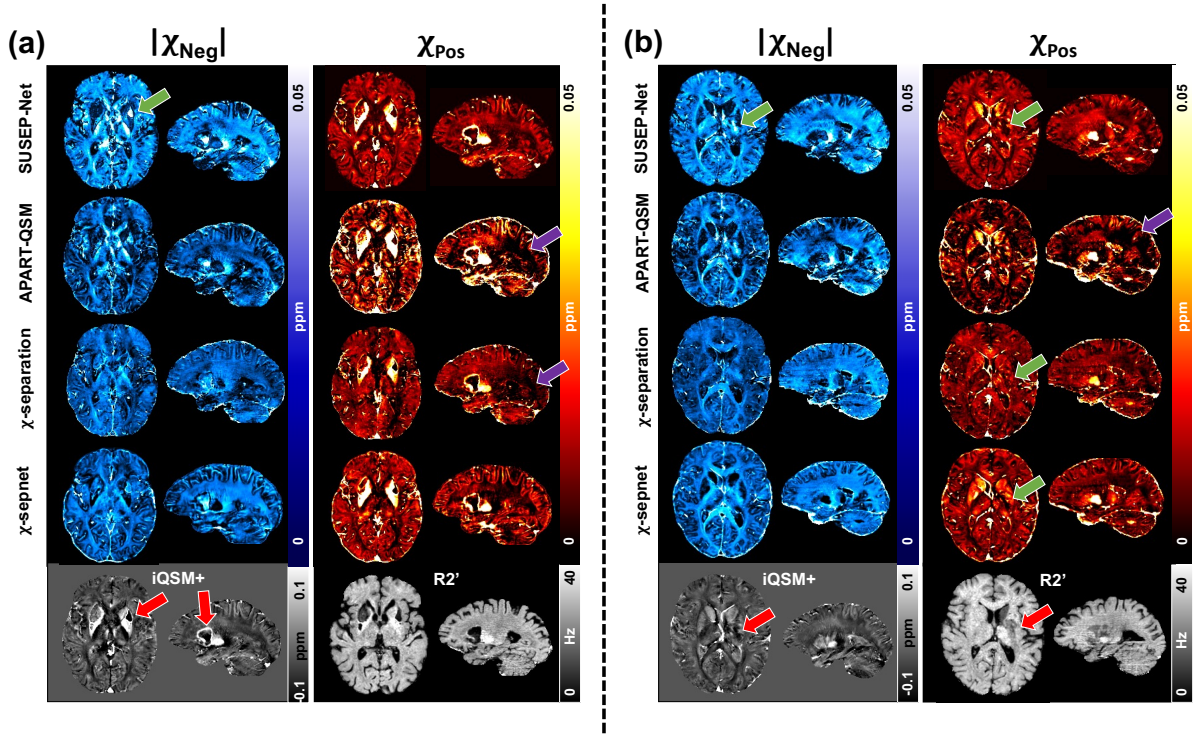

**Figure 8.** Susceptibility source separation results from (a) the CO poisoning and (b) MOGAD patients using different reconstruction methods. Red arrows point to the abnormal lesions in the QSM images, purple arrows point to the artifacts in traditional APART-QSM and χ-separation methods, while green arrows highlight the lesions apparent in SUSEP-Net reconstructions and less obvious in the other methods.

## 5. DISCUSSION AND CONCLUSION

In this study, we proposed a novel method, namely SUSEP-Net, for susceptibility source separation from the input local field, R2', and QSM images. The primary backbone of the proposed SUSEP-Net was constructed based on a dual-branch deep neural network, with one branch responsible for paramagnetic susceptibility map reconstruction and the other for diamagnetic susceptibility image. Extensive experiments were conducted to compare the performance of SUSEP-Net with three state-of-the-art susceptibility source separation methods (i.e., APART-QSM, χ-separation, and deep learning χ-sepnet) on simulated and *in vivo* healthy and pathological brains. The results showed that SUSEP-Net consistently led to better numerical results compared with other QSM separation methods in simulation studies, and demonstrated better image contrasts on the *in vivo* images from patients with CO poisoning and MOGAD. In addition, we also validated the generalization capability and accuracy of SUSEP-Net and the effectiveness of the proposed purely simulated training dataset with a specifically designed agarose gel phantom.

The proposed SUSEP-Net was trained based on a pure simulation dataset generated from APART-QSM images and purely synthetic high-intensity lesions. The benefit of this simulation-supervised strategy is that we can simulate a sufficient number of training pairs from only limited subjects, and more importantly, it ensures the training inputs and training labels satisfy exact underpinning physics. Through this framework, we expected that the SUSEP-Net could learn the underpinning physical model instead of simply imitating traditional algorithms to be more generalizable, which could be partially demonstrated by the improved image contrasts of SUSEP-Net against APART-QSM on the patients with pathological conditions.

The proposed simulation-supervised training makes it possible to get high-intensity susceptibility "lesions" involved in the network training, thereby improving the performance of SUSEP-Net on pathological brains, as confirmed in the corresponding experiments on both simulated and *in vivo* pathological brains. In addition, this simulation training is also beneficial to improve the deep learning-based methods' generalization capabilities on out-of-distribution data, as demonstrated in the phantom experiment. For example, the $\chi$-sepnet trained on our proposed pure simulated data achieved better qualitative and quantitative results than the original $\chi$-sepnet (Kim et al., 2025) trained with *in vivo* acquired data (as inputs) and the training labels calculated using the traditional $\chi$-separation method (Shin et al., 2021).

Apart from the dual-branch U-net design, another major difference between the proposed SUSEP-Net and the recently proposed U-net-based $\chi$-sepnet was the contrastive learning strategy designed in this work, which was inspired by recent contrastive representation learning (Chen et al., 2020; Chuang et al., 2020; Cole et al., 2022; Oord et al., 2018) and prompt learning (Zhou et al., 2022a; Zhou et al., 2022b) related works. It was designed to explicitly impose similarity-based constraints between the branch-specific guidance features calculated from the QSM input and the latent features that would be used for susceptibility source separation. Specifically, our design was to make the guidance and task-oriented latent features in the same branch more similar, while keeping the different branches more distinct, and the ablation study confirmed the effectiveness of the contrastive learning strategy for improving SUSEP-Net's performance.

Despite these advantages and improvements, the training of the proposed SUSEP-Net still leveraged an existing algorithm (e.g., APART-QSM) to calculate the training labels ($\chi_{pos}$ and $\chi_{neg}$) and the corresponding magnitude decay kernels (the A maps) for the simulation training strategy. Future works will investigate unsupervised algorithms to eliminate the dependencies of traditional methods for generating training labels. In addition, the computational cost of SUSEP-Net (17.96 seconds for a subject of size $224 \times 224 \times 128$) is also greater than the simple U-net-based methods (14.31 seconds) due to the complicated network design, and we should develop more light-weight networks for susceptibility source separation in the future.

**ACKNOWLEDGMENTS**

This work was supported by the National Natural Science Foundation of China under Grant No. 62301616 and 62301352, and the Natural Science Foundation of Hunan under Grant No. 2024JJ6530, Hunan Provincial Science and Technology Program (NO.2021RC4008), and the High Performance Computing Center of Central South University. HS thanks the support from the Australia Research Council (DE20101297 and DP230101628). GBP acknowledges support from the Canadian Institutes for Health Research (FDN 143290) and the Campus Alberta Innovates Chair. YG thanks Prof. Hongjiang Wei for the helps and instructions on APART-QSM codes and results.

**Data And Code Availability Statements**

The data are available upon request due to privacy/ethical restrictions. Source codes and trained networks are available at: https://github.com/YangGaoUQ/SUSEP-Net